\documentclass[aps,showpacs,twocolumn]{revtex4}  

\usepackage{float}

\usepackage{amsmath}
\usepackage{mathrsfs}
\usepackage{amssymb}
\usepackage{graphicx}
\usepackage{bm}
\usepackage{color}
\usepackage{times}
\usepackage{hyperref}
\usepackage{textpos}

\definecolor{clr}{rgb}{0,0.6,0.6}

\begin{document}

\title{Absolute frequency measurement of the 7s$^2$ $^1$S$_0$ $-$ 7s7p $^{1}$P$_1$ transition in $^{225}$Ra}

\author{B. Santra\footnote{present address: Research Center OPTIMAS, Technische Universit\"at Kaiserslautern, 67663 Kaiserslautern, Germany; bodhaditya@physik.uni-kl.de}}

\author{U. Dammalapati\footnote{present address: Department of Physics, The University of Dodoma, Dodoma, Tanzania}}
  
\author{A. Groot}

\author{K. Jungmann}

\author{L. Willmann}

  
\affiliation{ Van Swinderen Institute, Faculty of Mathematics and Natural Sciences, University of Groningen, Zernikelaan 25, 9747 AA Groningen, The Netherlands }  

\date{\today}

\begin{abstract}
Transition frequencies  were determined for transitions in Ra in an atomic 
beam and for reference lines in Te$_2$ molecules in a vapor cell. The absolute 
frequencies were calibrated against a GPS stabilized Rb-clock by means of 
an optical frequency comb. The 7s$^2\,^1$S$_0$(F = 1/2)-7s7p$\,^1$P$_1$(F = 3/2) 
transition in $^{225}$Ra was determined to be $621\,042\,124(2)\,$MHz. The 
measurements provide input for designing efficient and robust laser cooling 
of Ra atoms in preparation of a search for a permanent electric dipole moment 
in Ra isotopes.
\end{abstract}


\maketitle

\begin{figure}[ht]
\includegraphics[trim=2.7cm 1.75cm 2.0cm 1.9cm, clip=true, width=0.29\textheight, angle=0]{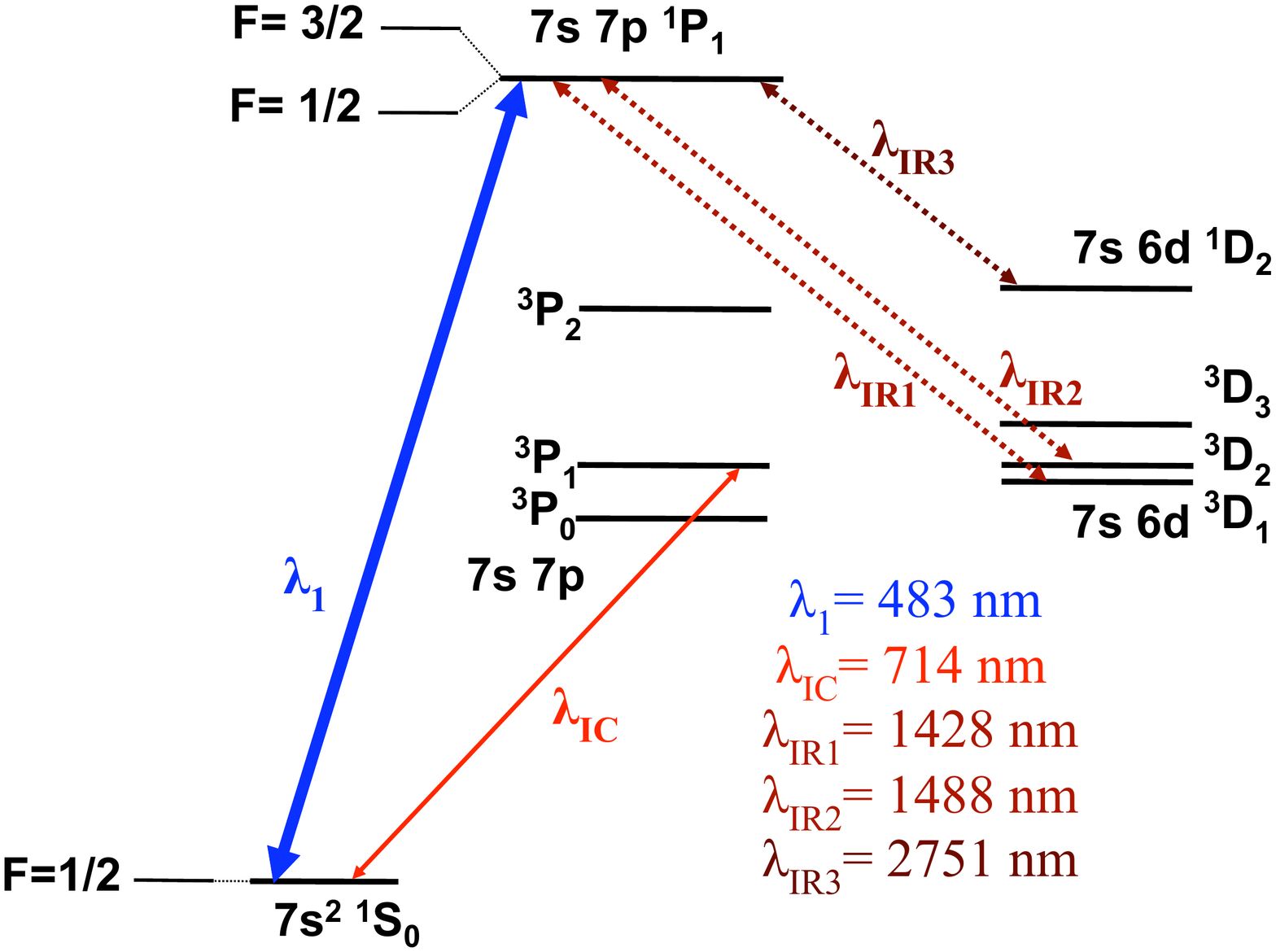}
\caption{(Color online) Lowest S, P and D states in atomic $^{225}$Ra. Transitions relevant for an EDM search are indicated. Dotted lines represent dominant leak channels for laser cooling via the 7s$^2\,^1$S$_0$-7s7p $^1$P$_1$ transition.}
\label{RaLevelscheme_483nm}
\end{figure}

Radium (Ra) is the heaviest alkaline earth metal and it offers unique possibilities for measuring parity and time reversal symmetry violation. The particular atomic and nuclear~\cite{Flambaum:1999eu,Engel:2000jk, Engel:2003xy, Dzuba:2000lq, Dobaczewski:2005qv,BieroN:2009ty, Dzuba:2007eu, BieroN:2007fv} structure in Ra isotopes cause the largest enhancement for permanent electric dipole moments (EDMs)~\cite{Scielzo:2006rt} in any atom. This arises from the close proximity of the 7s7p$\,^3$P$_1$ and 7s6d\,$^3$D$_2$ states~\cite{Flambaum:1999eu}. The exploitation of the enhancement from this $5\,$cm$^{-1}$ separation requires precise knowledge of Ra atomic properties such as the absolute frequencies of transitions that are relevant for laser cooling and state manipulation (see Fig.~\ref{RaLevelscheme_483nm}). Many isotopes of Ra are available from radioactive sources such as $^{229}$Th \cite{Willmann:2012jk,Jungmann:2003qv, Jungmann:2006,Guest:2007hb}, or at online isotope production facilities such as ISOLDE, CERN, Switzerland~\cite{Ahmad:1983jk,Wendt:1987}.

A sensitive search for EDMs requires efficient collection of the atoms in an optical trap because of the low abundance of Ra isotopes. A strategy for efficient laser cooling and trapping has been developed with the chemical homologue barium (Ba). Exploiting the strong 6s$^2\,^1$S$_0$-6s6p$\,^1$P$_1$ transition ~\cite{Dammalapati:2009, De:2009nx} resulted in an efficiency of $\sim$1$\%$ for slowing and capturing Ba from an atomic beam, whereas capture efficiencies of below 10$^{-6}$ were reported for Ra when using the weak intercombination transition 7s$^2\,^1$S$_0$-7s7p$\,^3$P$_1$\cite{Guest:2007hb}. 

The optical spectrum of Ra was first studied by Rasmussen~\cite{Rasmussen:1934}. This identified Ra as an alkaline earth metal. Hyperfine splittings and isotope shifts were determined for the 7s$^2\,^1$S$_0$-7s7p$\,^{1, 3}$P$_1$ transitions by collinear laser spectroscopy~\cite{Ahmad:1983jk, Wendt:1987} with intense Ra ion beams at ISOLDE, lifetimes of low lying states were determined in ANL, USA~\cite{Scielzo:2006rt,Trimble:2009}; no absolute frequencies are quoted. 
 
Here we report on a laser spectroscopy measurement of the frequency of the strong 7s$^2\,^1$S$_0$-7s7p$\,^1$P$_1$ transition in $^{225}$Ra in an atomic beam. The laser frequency was recorded with an optical frequency comb. A set of reference lines in molecular tellurium ($^{130}$Te$_2$) were measured simultaneously. An uncertainty of about 1 MHz is achieved. This is sufficiently small for efficient laser cooling on that transition compared to the natural linewidth of 30 MHz~\cite{Dzuba:2000lq}.


Doppler-free saturated absorption spectroscopy of $^{130}$Te$_2$ provides a reliable secondary frequency standard over a wide range in the visible spectrum. The absence of nuclear spin results in a spectrum without hyperfine structure. Many of the linear absorption lines are listed in the$^{130}$Te$_2$ atlas ~\cite{TelluriumAtlas} which reports the analysis of Fourier transform spectroscopy in the wavelength range of 450 nm to 600 nm. Several lines have been independently calibrated to MHz accuracy in interferometric measurements for particular experiments, e.g. for 1S-2S transitions in hydrogen, deuterium, positronium and muonium~\cite{Boshier:1989ty} or strong lines around 500 nm~\cite{Gillaspy:1991sf}. Deviations of up to 0.003 cm$^{-1}$ (100 MHz) from the values in reference \cite{TelluriumAtlas} have been found. This makes independent calibration of individual reference lines indispensable. The strongest Te$_2$ line in the vicinity of the 7s$^2\,^1$S$_0$(F = 1/2)-7s7p$\,^1$P$_1$(F = 3/2)transition in $^{225}$Ra is at 20\,715.477\,7 cm$^{-1}$ (Line 2004 in \cite{TelluriumAtlas}). 

\begin{figure}[ht]
\centering
\includegraphics[trim=2.3cm 1.0cm 4.9cm 1.2cm, clip=true, width=0.36\textheight, angle=0]{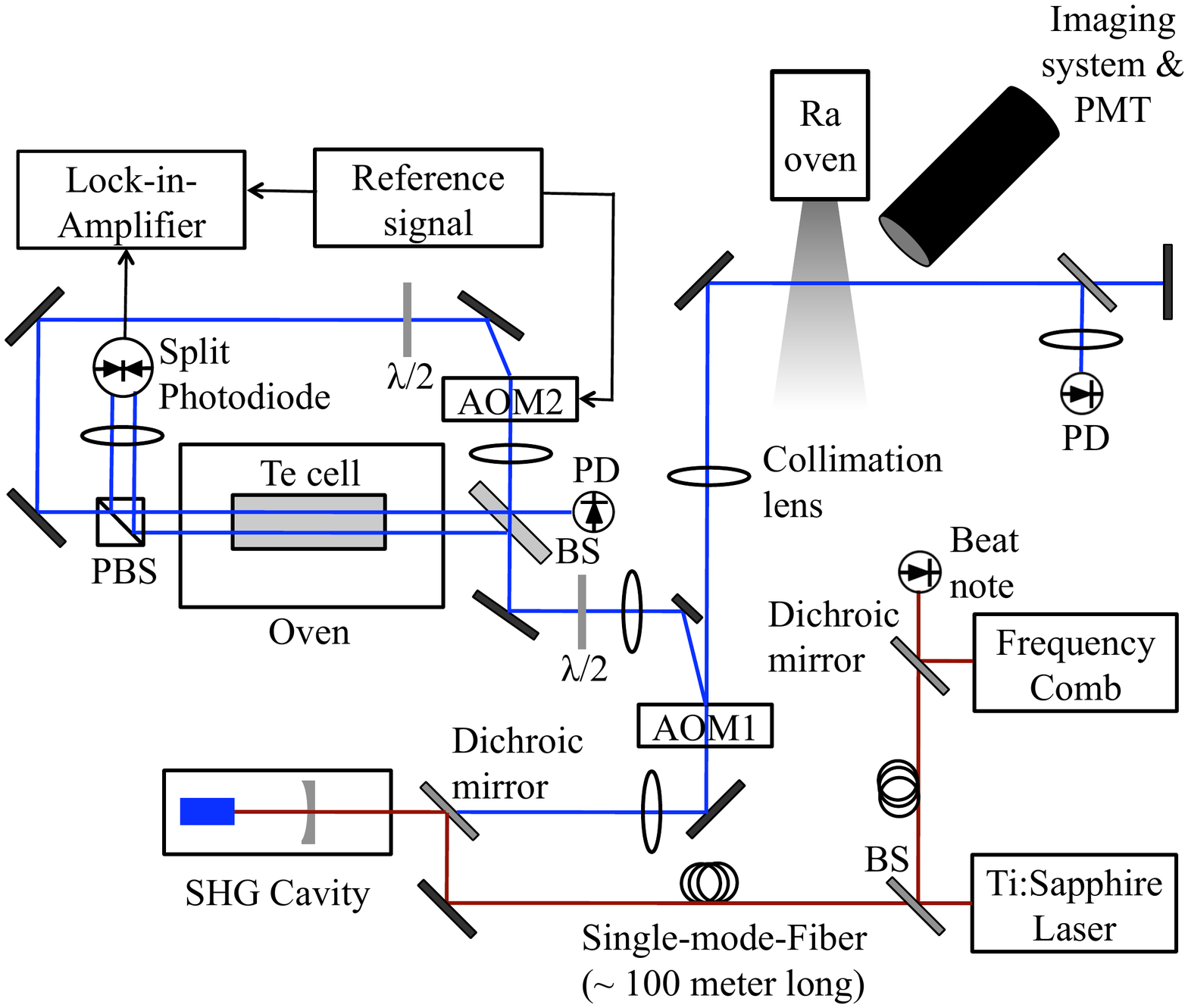}
\caption{(Color online) Schematic setup to measure the 7s$^2\,^1$S$_0$-7s7p$\,^1$P$_1$ transition frequency in $^{225}$Ra and saturated absorption in $^{130}$Te$_2$. Light from  a Ti:sapphire laser is frequency doubled to obtain light at wavelength $\lambda_{1}$. This beam is split in an acousto optical modulator AOM1 to obtain light for saturated absorption in Te$_2$ and fluorescence spectroscopy of  $^{225}$Ra. IR light is overlaped on a photodiode with light from an optical frequency comb to measure its absolute frequency.}
\label{TeSatAbs_Setup_publication}
\end{figure}

\begin{figure}[hb]
\centering
\includegraphics[trim=0.1cm 2.0cm 0.1cm 0.7cm, clip=true,width=0.22\textwidth,angle=90]{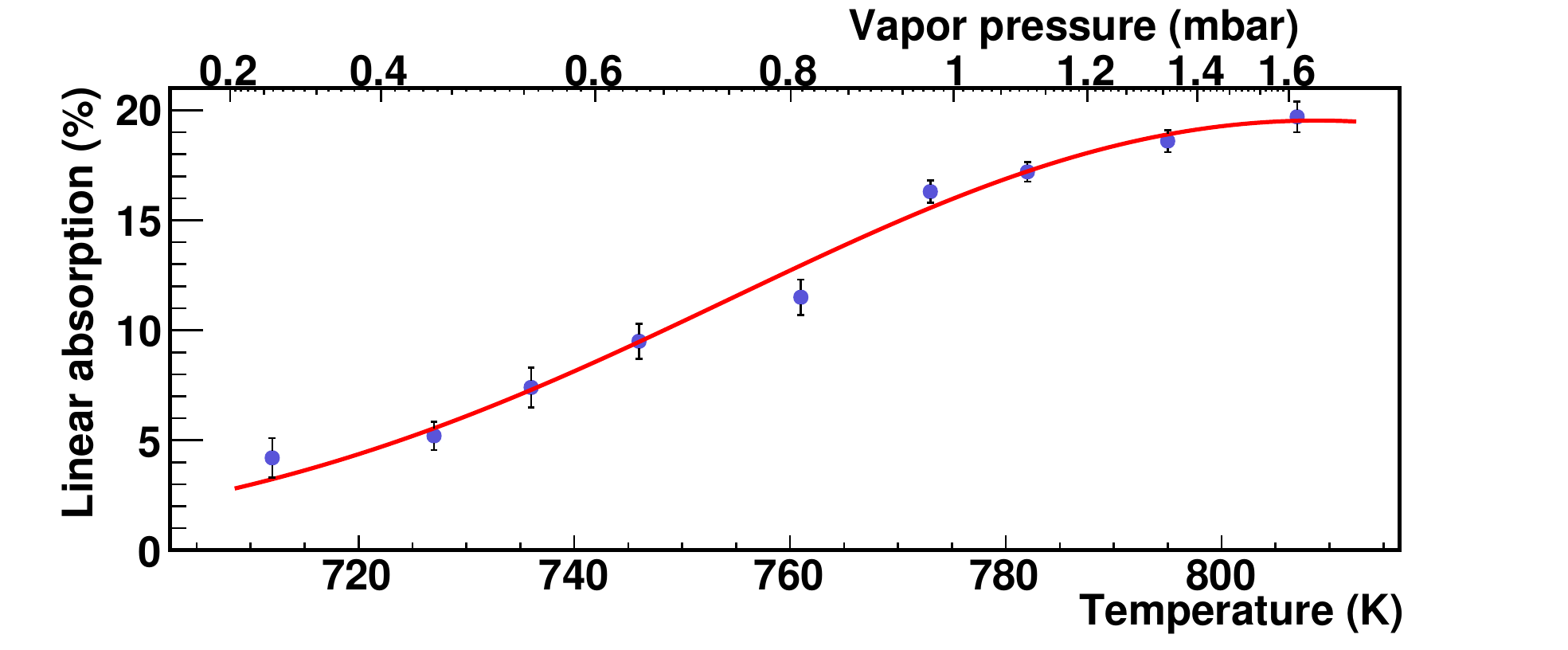}
\caption{(Color online) Teperature dependence of linear absorption for the $^{130}$Te$_2$ absorption line which corresponds to the Te$\#$3 line at 20 715.59 cm$^{-1}$. The data are fitted taking into account the temperature dependence of the vapor pressure of $^{130}$Te$_2$ from \cite{Machol:1958qv}.}
\label{TeAbs_AbsoluteTemp}
\end{figure}

\begin{figure}[ht]
\centering
\includegraphics[trim=6.2cm 11.7cm 5.1cm 10.25cm, clip=true, width=0.47\textwidth,angle=0]{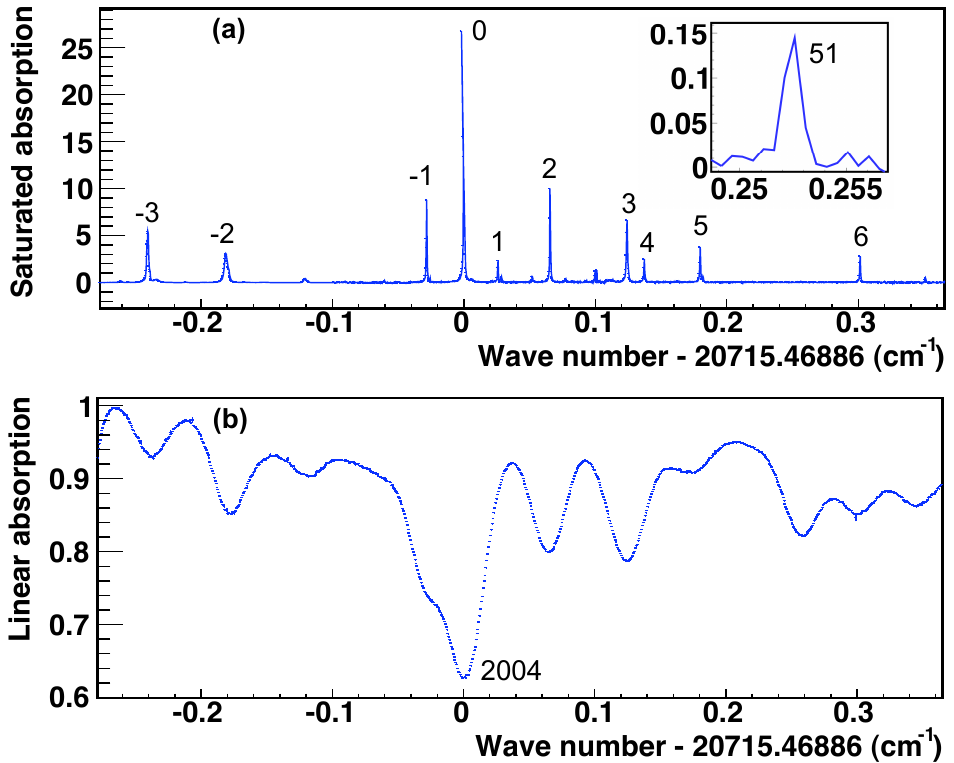}
\caption{(Color online) Saturated (a) and linear (b) absorption in  $^{130}$Te$_2$  covering absorption line 2004 of reference~\cite{TelluriumAtlas}. The line Te$\#$0 is calibrated to be centered at frequency 621\,034\,132.9(1.5) MHz or 20\,715.468\,86(5) cm$^{-1}$. {\it Inset}: The line labelled 51 at 20\,715.721\,48(5) cm$^{-1}$ is nearest to the 7s$^2$ $^1$S$_0$(F = 1/2) - 7s7p$^1$P$_1$(F$'$ = 3/2) transition in $^{225}$Ra.}
\label{Te_2004_thesis}
\end{figure}

\begin{figure}[ht]
\centering
\includegraphics[trim=6.0cm 11.1cm 5.3cm 10.6cm, clip=true, width=0.47\textwidth,angle=0]{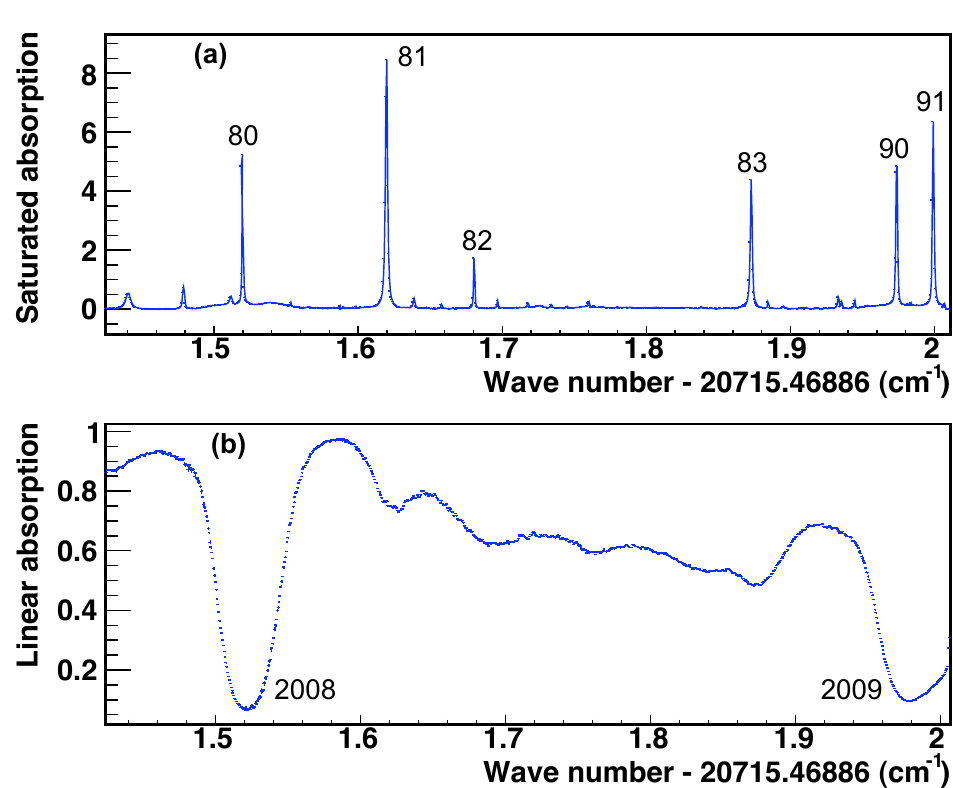}
\caption{(Color online) Saturated (a) and linear (b) absorption in $^{130}$Te$_2$ covering lines 2008 and 2009 in the $^{130}$Te$_2$ atlas~\cite{TelluriumAtlas}.}
\label{Te_2008_2009_thesis}
\end{figure}

Isotopically pure  $^{130}$Te$_2$ contained in a quartz cell of dimensions (100 mm $\times$  25 mm) is heated to temperatures between 700~K and 800~K for sufficient vapor density. The temperature is maintained by a  Watlow heater element of inner diameter 50~mm and length 150~mm. It is  monitored by a K-type thermocouple placed next to the cell. 
Light for the transitions in $^{225}$Ra and $^{130}$Te$_2$ is obtained through second harmonic generation (SHG) by feeding light from a Ti:Sapphire laser (Coherent MBR110) at wavelength 966~nm into a KNbO$_3$ crystal inside a linear enhancement cavity (see Fig. \ref{TeSatAbs_Setup_publication}). The light at wavelength 483~nm 
is focused into an acousto-optic modulator (AOM1, MT350, AA Optoelectronics) operated at frequency f$_{\mathrm{AOM1}}$ = 450 MHz. It provides a frequency offset, which bridges the separation between the nearest line in $^{130}$Te$_2$ and the transition in $^{225}$Ra. The 0$^{\mathrm{th}}$-order beam from AOM1 serves for saturated absorption in Te$_2$ and the minus first-order beam is used for spectroscopy on $^{225}$Ra. For the latter the light is overlapped with an effusive $^{225}$Ra atomic beam from an oven filled with 30$\,\mu$g of $^{229}$Th (half-life 7340~y), coresponding to source strength 10 $\mu$Ci. The oven is contained inside a UHV vacuum chamber with residual gas pressure below $10^{-9}$ mbar. Typically $^{225}$Ra atoms were accumulated for one week inside the cavity of the oven,
which when heated to temperature 900 K produced a flux of 10$^6$ s$^{-1}$ cm$^{-2}$ for about
one hour. Optical windows for laser beam access enable the alignment of the laser beam orthogonal to the atomic Ra beam. Fluorescence light is collected through an interference filter under right angle to both beams with a photomultiplier tube (PMT).

\begin{table}[ht]
\begin{center}
\caption{Wavenumbers, frequencies and relative strengths of saturated absorption lines in $^{130}$Te$_2$. The uncertainty of the wavenumber measurement for line Te$\#$0 is 0.000\,05 cm$^{-1}$ corresponding to 1.5 MHz for the optical frequency. We quote both frequencies and wavenumbers for the transitions to enable comparison with the  $^{130}$Te$_2$ atlas~\cite{TelluriumAtlas} and previous measurements~\cite{McIntyre:1990jk, Gillaspy:1991sf, Barr:1985xy}}
\begin{tabular}{cccccc}
\hline
\hline
  Saturated & Absorption & Measured  &\hspace{0 mm} & Measured &Relative \\ 
  absorption & line in & Wave number  && Frequency &strength\\ 
   line & Te Atlas & (cm$^{-1}$) && (MHz) &\\ 
  \hline
  -3 &  & 20\,715.222\,52 && 621\,026\,747.7(1.0) & 0.23\\

  -1 & & 20\,715.440\,56 && 621\,033\,284.4(1.2) & 0.37\\

  0 & $\#$ 2004 & 20\,715.468\,86 && 621\,034\,132.9(1.5) & 1.00\\

  1 & & 20\,715.494\,32 && 621\,034\,896.2(1.6) &  0.09\\

  2 & & 20\,715.534\,21 && 621\,036\,092.1(0.7) & 0.34\\

  3 & & 20\,715.593\,74 && 621\,037\,876.7(1.6) & 0.19\\

  4 & & 20\,715.607\,49 && 621\,038\,288.8(1.5) & 0.09\\

  5 & & 20\,715.651\,12 && 621\,039\,596.9(1.6) & 0.14\\

  51 & & 20\,715.721\,48 && 621\,041\,706.3(1.7) & 0.008\\

  6 & & 20\,715.770\,29 && 621\,043\,169.4(1.5) & 0.16\\

  80 & $\#$ 2008 & 20\,716.989\,05 && 621\,079\,707.1(0.2) & 0.12\\

  81 & & 20\,717.088\,18 && 621\,082\,678.8(0.7) & 0.21\\

  82 & & 20\,717.150\,13 && 621\,084\,536.0(0.8) & 0.03\\

  83 & & 20\,717.341\,67 && 621\,090\,278.2(0.4) & 0.06\\

  90 & $\#$ 2009 & 20\,717.436\,66 && 621\,093\,126.1(0.3) & 0.15\\

  91 & & 20\,717.460\,39 && 621\,093\,837.3(0.4) & 0.16\\
  \hline
   \hline
\end{tabular}
\label{Frequencies_Strengths_Te}
\end{center}
\end{table}

The light for $^{130}$Te$_2$ spectroscopy is passed through a thick beamplitter plate (BS) to provide two probe beams and one pump beam. The parallel probe beams are passed though the $^{130}$Te$_2$ cell. The pump beam at intensity 400(50) $\mu$W/mm$^2$ is focused into an acousto-optic modulator (AOM2) operated at f$_{\mathrm{AOM2}}$ = 60 MHz. The driving rf power is chopped at f$_{\mathrm{mod}}$ $\approx$ 13 kHz to provide amplitude modulation of the beam. This light is overlapped on a polarizing beam splitter cube (PBS) with one of the probe beams in the $^{130}$Te$_2$ cell. The difference in absorption for both probe beams is measured on a balanced photo detector (PD). This signal is demodulated in a lock-in amplifier to obtain the saturated absorption signal. 
The vapor pressure in the $^{130}$Te$_2$ cell is monitored by linear absorption that corresponds to the Te$\#$3 saturated absorption line (see Fig.~\ref{TeAbs_AbsoluteTemp}). A model from~\cite{Barwood:1991xy, Machol:1958qv} fits the data with $\chi^2/5$ =  1.146.

\begin{figure}[h]
\centering
\includegraphics[trim=6.62cm 11.5cm 6.55cm 9.95cm, clip=true, width=0.5\textwidth,angle=0]{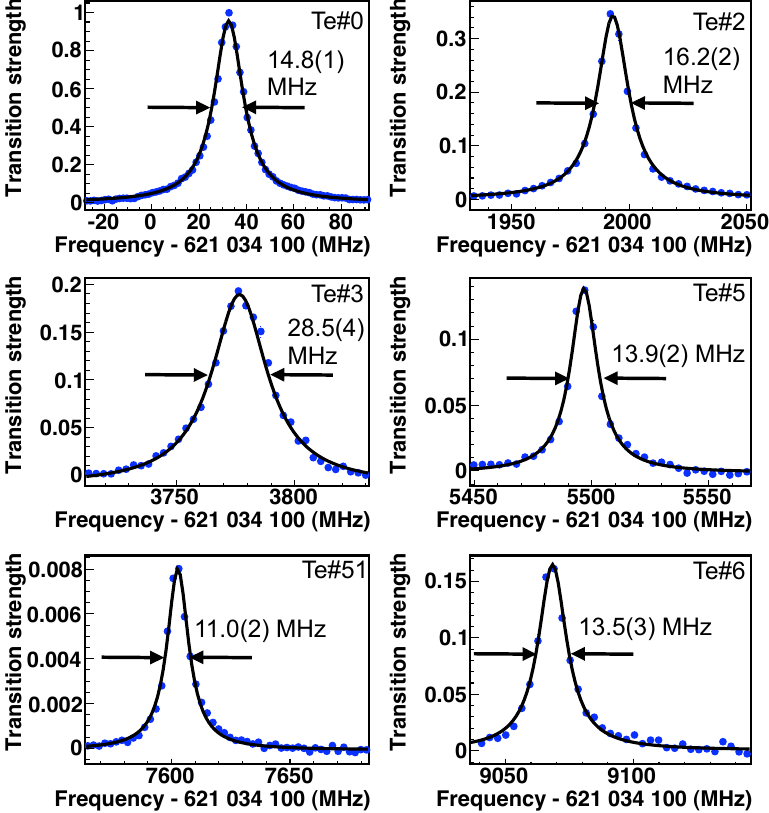}
\caption{(Color online) Six individual saturated absorption lines in $^{130}$Te$_2$ as a function of absolute frequency. The full width at half maximum for all lines is determined by fitting a Lorentzian function to the data.}
\label{Te_Lines_detail_thesis_BS}
\end{figure}

A fraction of the infrared (IR) light from the Ti:Sapphire laser overlaps with light from the frequency comb (Menlo Systems FC1500-250) on a photodiode to measure their beatnote at frequency 
$f_{BN}$. The frequency $f_{Te}$  of the transition in $^{130}$Te$_2$ is
\begin{equation}
\label{TeFreq}
f_{Te}= 2 \left(mf_{rep}+2f_{CEO}+f_{BN}\right) +n_{1} f_{1}+ \dfrac{n_{2} f_{2}}{2}.
\end{equation}
Here $m$ is the mode number, $f_{rep}$ is the repetition rate and $f_{CEO}$ = 20 000 000 Hz is the carrier envelope offset frequency of the comb. The long term accuracy of the frequency comb derives from a Global Positioning System (GPS) stabilized Rb clock to better than 10$^{-11}$ \cite{Ludlow:2008ty}. This corresponds to a frequency uncertainty of 0.006~MHz for our measurments in $^{130}$Te$_2$.
The mode number $m$ is determined with a wavelength meter (HighFinesse \AA ngstrom WS6 VIS) that was calibrated with light at sufficiently known frequency. 
$n_{1}$ and $n_{2}$ are the sideband diffraction orders from AOM1 and AOM2, while $f_{1}$ and $f_{2}$ are the respective operating frequencies. Fig.~\ref{Te_2004_thesis} and Fig.~\ref{Te_2008_2009_thesis} display saturated (top) and linear absorption (bottom) spectra in $^{130}$Te$_2$. 
The repetition rate $f_{rep}$, the offset frequency $f_{CEO}$ and the operation parameters for AOM1 and AOM2 were all kept constant while the IR light frequency was scanned across the transitions. Six of the recorded lines are displayed in Fig.~\ref{Te_Lines_detail_thesis_BS}. The measured saturated absorption lines are fitted with Lorentzian line shapes, 
\begin{equation}
\label{lorfit}
L(\omega) = \dfrac{1}{2\pi}\dfrac{\Gamma}{(\omega - \omega_0)^2 + \Gamma^2/4} \ ,
\end{equation}
where $\omega_0$ is the center frequency and $\Gamma$ is the width of the profile.

Linear absorption of the pump beam was measured by transmitting light through the $^{130}$Te$_2$ cell for vapor pressures ranging from 0.2 mbar to 1.6 mbar. Saturated absorption was measured at cell temperature 804(5)~K and Te$_2$ vapor pressure $\sim$1.6~mbar. For all individual saturated absorption lines frequency uncertainties due to the fitting procedure are of order 0.1 MHz. 
The linewidths of the transitions is between 11 and 29 MHz while the signal strengths varied by a factor 125 in the range from line Te$\#$0 (Fig.~\ref{Te_2004_thesis}) to line Te$\#$51 (Fig.~\ref{Te_2004_thesis}({\it Inset})). The frequencies and wavenumbers for the saturated absorption lines are listed in Table~\ref{Frequencies_Strengths_Te}. 

The pressure shift~\cite{Demtroder:2003,Brebrick:1968jk} of similar lines in $^{130}$Te$_2$ has been determined previously to be 1 MHz/mbar ~\cite{Barr:1985xy, Barwood:1991xy, McIntyre:1987jk, McIntyre:1990jk}. During the measurements the maximum observed fluctuation of the $^{130}$Te$_2$ cell temperature was $\pm$ 5 K. This limits the frequency uncertainty due to pressure shifts for the lines measured in $^{130}$Te$_2$ to below 0.2 MHz. 

\begin{figure}[ht!]
\centering
\includegraphics[trim=0.1cm 2.1cm 0.1cm 0.4cm, clip=true, width=0.25\textwidth,angle=90]{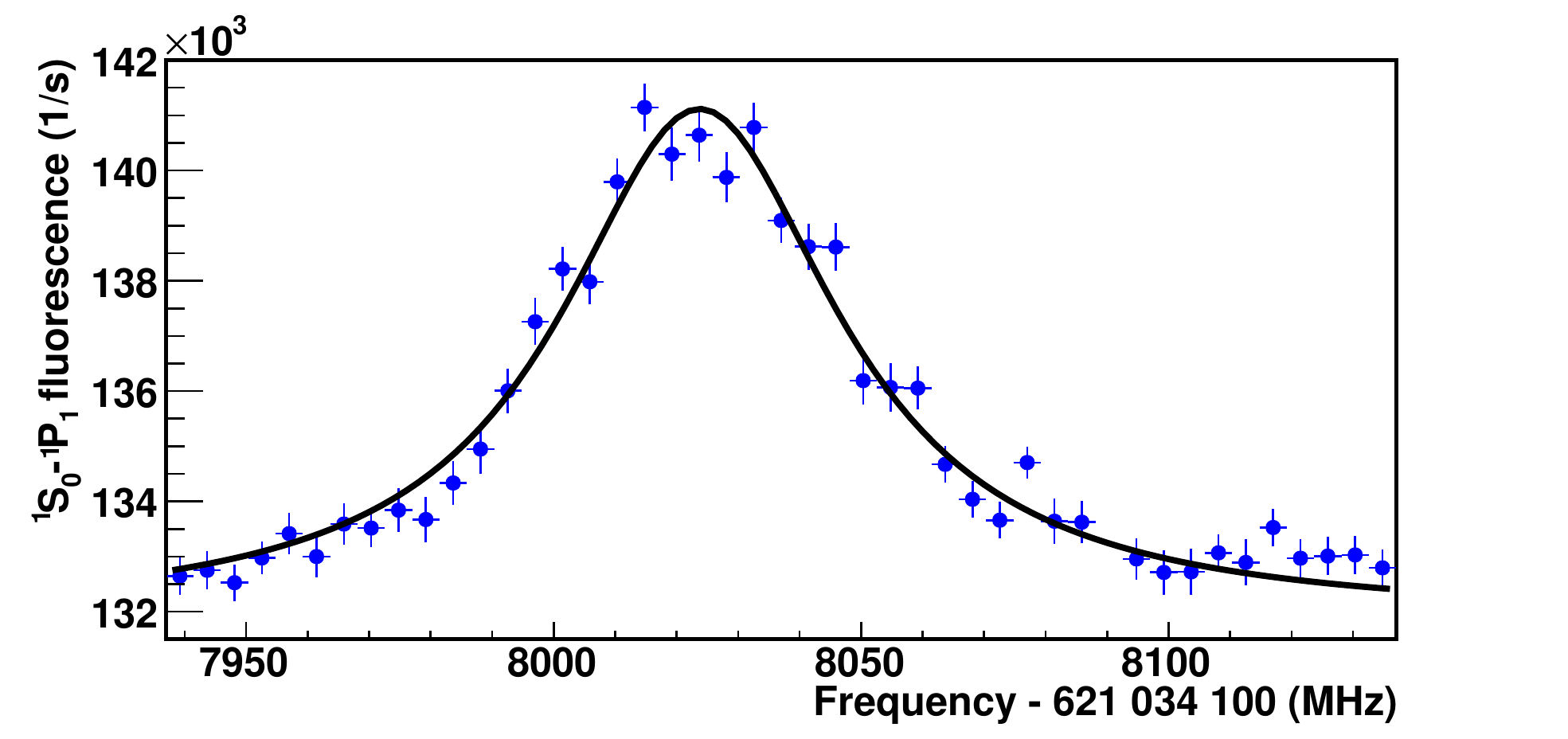}
\caption{(Color online) Doppler-free laser induced fluorescence from the $^1$S$_0$(F = 1/2) $-$ $^1$P$_1$(F$'$ = 3/2) transition in $^{225}$Ra. The data are averaged over five individual scans of duration 50~s. The line is fitted with a Lorentzian function ( $\chi^2/41$ = 1.256).}
\label{Ra_483nm_13Apr2011_thesis_1}
\end{figure}

\begin{table}[h]
\begin{center}
\caption{The frequency of the $^1$S$_0-^1$P$_1$ transition in $^{226}$Ra has been determined in different experiments. In this work the frequency of the $^1$S$_0$(F = 1/2)-$^1$P$_1$(F$'$ = 3/2) transition in $^{225}$Ra was measured. The frequency of this transition in $^{226}$Ra is obtained by exploiting this measurement, the known isotope shift in the $^1$S$_0-^1$P$_1$ transition $f_{^{225}Ra}$ $-$ $f_{^{226}Ra}$ $ = $ 2236 (15) MHz~\cite{Wendt:1987}, and the magnetic dipole interaction constant A ($^1$P$_1$) $ = $ 2\,796.5(2.5) MHz of $^{225}$Ra~\cite{Ahmad:1983jk}.}
\begin{tabular}{ccclc}
\hline
\hline
 Isotope &\hspace{20 mm}& Transition & Frequency  & Experiment \\ 
 & & & (MHz)  & \\ 
  \hline
 \vspace{-4.0 mm}\\

$^{226}$Ra && $^1$S$_0$ $-$ $^1$P$_1$ & 621\,038\,489 (15) & This work\\

$^{226}$Ra && $^1$S$_0$ $-$ $^1$P$_1$ & 621\,038\,004 (180) & \cite{Trimble:2009}\\

$^{226}$Ra && $^1$S$_0$ $-$ $^1$P$_1$ & 621\,041\,362 (1 500) & \cite{Rasmussen:1934}\\
  \hline
   \hline
\end{tabular}
\label{Absolute_Frequency_Ra_483nm}
\end{center}
\end{table}


The absolute frequency of the $^1$S$_0$(F = 1/2)-$^1$P$_1$(F$'$ = 3/2) transition in $^{225}$Ra is measured with saturated absorption line Te$\#$51 in $^{130}$Te$_2$ as a reference. The transition frequency in Ra is obtained through 
\begin{equation}
f_{Ra} = f_{Te} + n_{1}f_{1}- \dfrac{n_{2}f_{2}}{2} + \Delta f_{TeRa} ,
\label{f_Ra_Te}
\end{equation}
where $\Delta f_{TeRa}$ is the observed frequency difference between the reference line and the $^{225}$Ra transition. We find the transition in  $^{225}$Ra (see Fig.~\ref{Ra_483nm_13Apr2011_thesis_1}) to be centered 418(1)~MHz below the reference. This yields an absolute frequency of $f_{^{225}Ra}$ = 621\,042\,124(2)~MHz, respectively 20\,715.735\,42(6) cm$^{-1}$. Together with the isotope shift in this transition of $f_{^{225}Ra}$ $-$ $f_{^{226}Ra}$ $ = $ 2\,236 (15) MHz~\cite{Wendt:1987} and the magnetic dipole interaction constant A($^1$P$_1$) = 2\,796.5(2.5)~MHz for $^{225}$Ra~\cite{Ahmad:1983jk}
we have the transition frequency $f_{^{226}Ra}$ = 621\,038\,489(15)~MHz 
for this transition in $^{226}$Ra (see Table~\ref{Absolute_Frequency_Ra_483nm}). 
The measurement here was performed with a retroreflected beam (see Fig.~\ref{TeSatAbs_Setup_publication}). The overlap angle of both beams was better than 1~mrad, corresponding to a separation $\leq$ 1~mm of the counter-propagating beams at distance 1~m from the interaction region. The beam alignment was optimized by minimizing the linewidth of the fluorescence signal. This causes residual first-order Doppler shift $\Delta_{Doppler}$ $\leq$ $2\,$ MHz for the thermal atomic beam in our experiment.


We exploited an offline atomic beam of $^{225}$Ra for Doppler-free laser induced spectroscopy on the $^1$S$_0$ $-$ $^1$P$_1$ transition. It is the main transition for efficient slowing of atoms to within the capture range of a magneto optical trap.  Our measurement of the transition frequency with uncertainty 2~MHz is an improvement by some two orders of magnitude over a previous measurement for the same transition in $^{226}$Ra. This together with the calibration of several lines in Te$_2$ molecules to accuracy MHz provides crucial input for the design of a sensitive search for an EDM in atomic Ra. 

This work was supported by  the Dutch Stichting voor Fundamenteel Onderzoek der Materie (FOM) under program 114 (TRI$\mu$P). BS acknowledges support from a Ubbo Emmius PhD scholarship from RUG.


\bibliography{bib-publication-BS}

\end{document}